\documentclass{appolb}
\usepackage{graphicx}
\usepackage{amsmath}

\begin{document}
\title{Proton-neutron random phase approximation studied 
by the Lipkin-Meshkov-Glick model in the SU(2)$\times$SU(2) basis%
\thanks{Presented at XXIII Nuclear Physics Workshop ``MARIE \& PIERRE CURIE'', Kazimierz Dolny, Poland 2016}%
}
\author{F. Minato
\address{Nuclear Data Center, Japan Atomic Energy Agency, Tokai, Japan}
}
\maketitle
\begin{abstract}
We study the proton-neutron RPA with an extended Lipikin-Meshkov-Glick model. We pay attention to the effect of correlated ground state and the case in which neutron and proton numbers are different. The effect of the correlated ground state are tested on the basis of quasi-boson approximation. We obtain the result that RPA excitation energies and transition strengths are in a good agreement with the exact solution up to a certain strength of the particle-particle interaction. However, the transition strength becomes worse if we consider the case in which neutron and proton numbers are different even at a weak particle-particle interaction. 
\end{abstract}

\PACS{21.00.00, 21.10.Re, 21.60.-n}

\section{Introduction}
The random phase approximation (RPA) is one of the useful approaches to describe a collective motion of nuclei and helps us to understand the basic mechanism of nuclear excitations. Its application to charge exchange reaction is widely used in calculations of neutrino-nucleus reactions \cite{Suzuki2013}, $\beta$-decay \cite{Minato2013} and isospin symmetry breaking \cite{Liang2010}. RPA is able to provide the basic physical insight of nuclear excitations by its simple picture of coherent 1 particle-1 hole (1p1h) excitations one hand, it doesn't describe coupling to more complicated states, like phonon coupling as well as multi-particle multi-hole states on the other hand. Therefore, several approaches beyond RPA have been also studied, for example, particle-vibration coupling \cite{Niu2014}, finite-rank separable approximation \cite{Severyukhin2014}, second RPA \cite{Papakonstantinou2014, Gambacurta2015} and Tamm-Dancoff-approximation (TDA) \cite{Minato2016}.

To take into account the higher order correlation beyond RPA, there is another approach which focuses on the ground state. When one derives the standard RPA, the RPA ground state, namely, the correlated ground state, is replaced with the Hartree-Fock (HF) one. This prescription omits a part of the multi-particle multi-hole effects. In this respect, several extensions of RPA to include the correlation have been studied. Renormalized RPA \cite{Rowe1968}, which considers renormalized single particle states invoked by a correlated nuclear ground state, is a leading example. While it is recognized that considering the ground state correlation is important to describe the nuclear collective vibration more practically than using HF basis, it is pointed out that it is not so significant in case of charge exchange reactions. This idea is based on the fact that proton and neutron have the different Fermi energies and occupy different shells. This might be true for heavy $N>Z$ nuclei. In fact, proton-neutron RPA calculation for $N>Z$ nuclei show almost the same result as proton-neutron TDA calculation, which implies the ground state correlation is weak enough. However, we should keep it in mind that RPA describing a specific transition doesn't take into account all the ground state correlation.

It is a good way to study RPA using with and without the correlated ground state by an exact solvable model in order to check the validity of the uncorrelated ground state. To this end, Lipkin-Meshkov-Glick (LMG) model \cite{Lipkin1965} can be a good tool because it enables us to compare the model with the exact solution in a simple model space. It has been widely used so far to validate various kinds of models with interests \cite{Gambacurta2006, Hagino2000, Hagino2000b, DinhDang2003}. To check the validity of RPA in case of charge exchange reactions, LMG model in SU(2)$\times$SU(2) basis was studied by Stoica's group \cite{Stoica1997, Stoica2001}. According to their results, RPA works well in case of a small nucleon number system, if the particle-particle interaction is weak enough relatively. They also considered the effect of correlated ground state up to the first order \cite{Stoica1998}. In this formalism, the ground and excited states of mother and daughter nuclei are first calculated with RPA and then consider the transition between them. They compared the transition strengths calculated by RPA on the basis of the correlated ground state with the exact solutions, and showed that RPA works reasonably well. Then one may think whether the same result can be obtained in case of proton-neutron RPA. It should be mentioned that reliability of RPA and quasi-particle RPA (QRPA) as well as renormalized QRPA for charge-exchange reactions has been also investigated in several different ways \cite{Civitarese1991, Delion2000, Stetcu2004, Civitarese2005}.

In this work, we present the effect of the correlated ground state characterized by phonon operator of proton-neutron RPA with the LMG model in SU(2)$\times$SU(2) basis. What is different from Ref. \cite{Stoica1998} is that charge exchange phonon creation operators are used to construct the excited and the correlated ground states. We particularly pay attention to nuclei with different neutron and proton numbers. Our formalism is based on the work of Ref. \cite{Stoica2001}, however they didn't investigate the effect of the correlated ground state. As shown in the next section, we obtain the different result from $N=Z$ nuclei in case of $N\ne Z$. A very similar work has been performed in case of SO(5) group \cite{Civitarese2005, Hirsch1997}, but the present work using SU(2)$\times$SU(2) will give another insight about the effect of the correlated ground state.

This paper organizes as follows. Sec. \ref{calculation} describes our formalism briefly. Sec. \ref{result} shows the result and compare RPA with the exact one and sec. \ref{summary} gives summary of this paper.

\section{Calculation}
\label{calculation}
Our model is almost the same as the work of Ref. \cite{Stoica2001}. However, we would like to describe only the key point briefly. We use the SU(2)$\times$SU(2) group algebra characterized by $T_+^{(1)}, T_-^{(1)}, T_z^{(1)}, T_+^{(2)}, T_-^{(2)}, T_z^{(2)}$ defined in \cite{Stoica2001}. Let's consider two levels each for proton and neutron. As defined in Ref. \cite{Stoica2001}, $p+(n+)$ and $p-(n-)$ are the symbols representing the higher and the lower levels of proton (neutron). The Hamiltonian considered in this work is
\begin{equation}
H=\epsilon(T_z(1)+T_z(2))+V_{pn}(T_+^{(1)}T_+^{(2)}+T_-^{(2)}T_-^{(1)})
+W_{pn}(T_+^{(1)}T_-^{(1)}+T_+^{(2)}T_-^{(2)}),
\label{hamil}
\end{equation}
where $\epsilon$ is the energy difference between the lower and higher levels of proton and neutron. The third and forth terms of Eq. \eqref{hamil} are the particle-particle and particle-hole interactions. To diagonalize the Hamiltonian, we consider the following basis as the set of the eigenvectors,
\begin{equation}
|\mu\rangle=|T^{(1)},T_z^{(1)}\rangle \otimes|T^{(2)},T_z^{(2)}\rangle,
\end{equation}
where the index $\mu$ stands for $\mu=(T_z^{(1)},T_z^{(2)})$. $T^{(1)}=N_n/2$ and $T^{(2)}=N_p/2$, where $N_n$ and $N_p$ are the neutron and proton numbers. The uncorrelated ground state is then given by $|0\rangle\equiv|T^{(1)},-T^{(1)}\rangle \otimes|T^{(2)},-T^{(2)}\rangle$. The Hamiltonian given in Eq. \eqref{hamil} can be exactly diagonalized by the linear combination,
\begin{equation}
|\Psi_i \rangle=\sum_{\mu}c_{\mu i}|\mu\rangle,
\end{equation}
where $i$ stands for eigenstates. The RPA formalism is also the same as Ref. \cite{Stoica2001}. To take into account the correlated ground state, we follow the same prescription as \cite{Rowe1968, Gambacurta2006, Stoica1998}. Up to the first order, it is given by
\begin{equation}
|RPA \rangle \sim N_0\left(1-\frac{1}{2N}\sqrt{\frac{\epsilon-\Omega}{\epsilon+\Omega}}\Theta^\dagger\Theta^\dagger\right)|0\rangle,
\label{RPAstate}
\end{equation}
where $N=N_n+N_p$, $\Omega>0$ is the eigenvalue of the RPA equation, and $N_0$ is the normalization factor satisfying $\langle RPA| RPA\rangle=1$. The second term of Eq. \eqref{RPAstate} takes into account 1 proton particle 1 neutron particle-1 proton hole 1 neutron hole $[\pi\nu(\pi\nu)^{-1}]$ configurations in addition to the 0 particle-0 hole configuration appearing in the first term \footnote{Eq. \eqref{RPAstate} also includes 2 proton particle-2 neutron hole and 2 neutron particle-2 proton hole configurations. However, they are not important because the Hamiltonian of Eq. \eqref{hamil} doesn't allow to form such a configuration in the ground state.}. We refer to results using Eq. \eqref{RPAstate} as RPA(corr.) in what follows. The transition strength for $\beta^-$ transition in RPA is then given by
\begin{equation}
\begin{split}
T_-&=|\langle 1|M^+|RPA \rangle|^2 \sim |\langle 1|M^-|0 \rangle|^2\\
\end{split}
\label{strength}
\end{equation}
where $|1\rangle=\Gamma^\dagger|RPA\rangle\sim\Gamma^\dagger|0\rangle$ and the transition operator $M^+$ is given in \cite{Stoica2001}. The phonon operator $\Gamma^\dagger$ is given by Eq. (8) of \cite{Stoica2001} in case of RPA, and the denominator of it is replaced by $\sqrt{\langle RPA|[\Theta^-,\Theta^+]|RPA \rangle}$ in case of RPA(corr.). The second and third equations of Eq. \eqref{strength} correspond to that of RPA(corr.) and RPA, respectively. Similarly, the transition strength for $\beta^+$ transition is given by
\begin{equation}
T_+=|\langle 1|M^-|RPA \rangle|^2 \sim |\langle 1|M^+|0 \rangle|^2.
\end{equation}

\section{Result}
\label{result}
\begin{figure}
\centering
\includegraphics[width=0.46\linewidth]{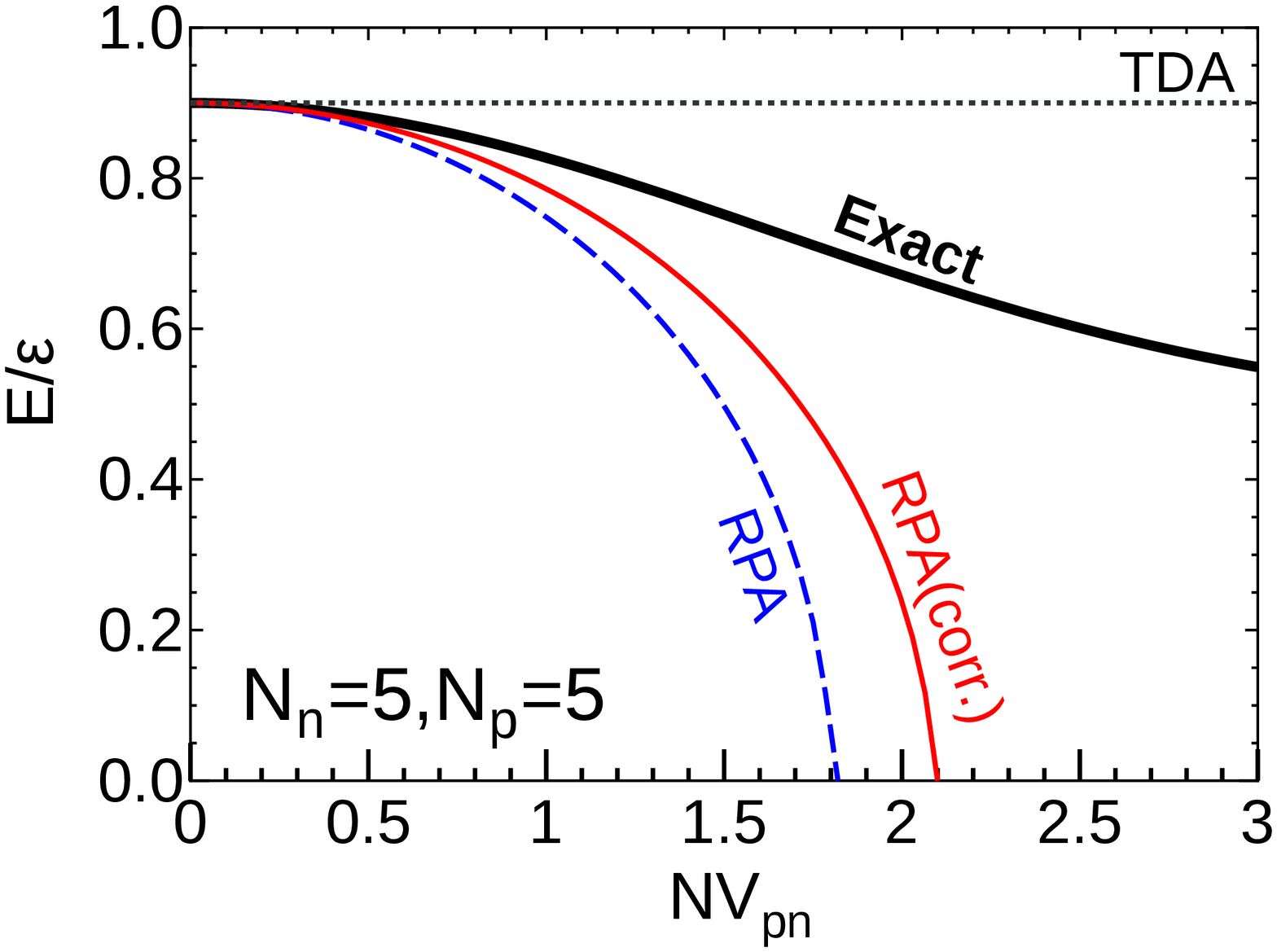}
\includegraphics[width=0.46\linewidth]{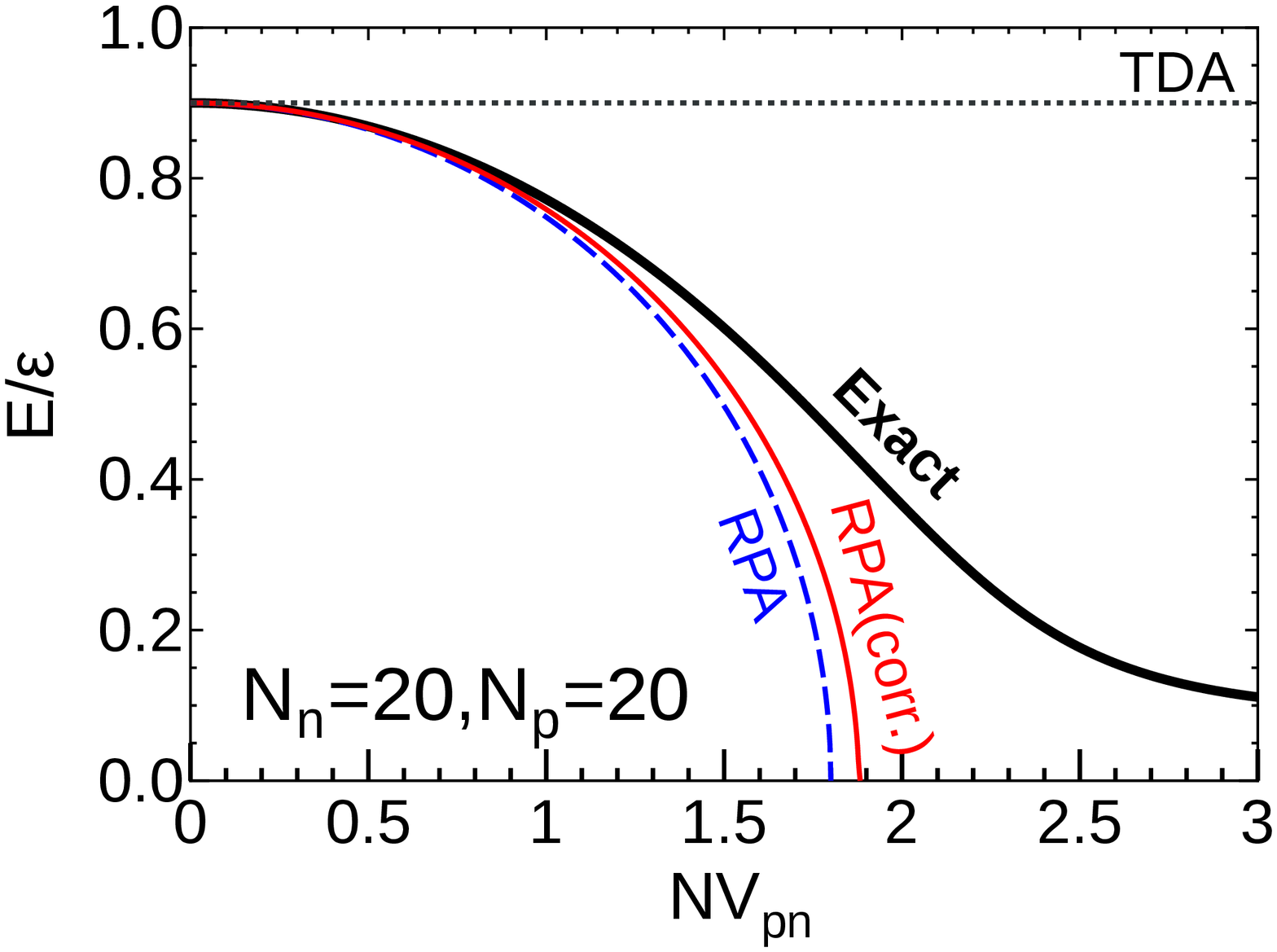}
\caption{Excitation energies in case of $N_n=N_p=5$ (left) and $N_n=N_p=20$ (right) as a function of $NV_{pn}$. The thick solid, thin solid, dashed, and dotted lines are the results for the exact, RPA(corr.), RPA, and TDA.}
\label{eNeqZ}
\end{figure}

First of all, we discuss in the case of which neutron and proton numbers are same. Figure \ref{eNeqZ} shows the excitation energy of $N_n=N_p=5$ (the left panel) and $N_n=N_p=20$ (the right panel). We set the model parameter of the particle-hole interaction as $NW_{pn}=-0.2$. We also compare our result with TDA which can be obtained in RPA by setting the backward amplitude $Y=0$. RPA and RPA(corr.) results show a similar curve to the exact one at a small $NV_{pn}$. At a certain large $NV_{pn}$ (critical point), both RPA and RPA(corr.) collapse due to the phase transition, however, RPA(corr.) has a larger critical point than RPA. RPA(corr.) result is closer to the exact one than RPA one both for $N_n=N_p=5$ and $N_n=N_p=20$, but the difference between them becomes smaller in case of $N_n=N_p=20$. Namely, the effect of the correlated ground state becomes not so significant for nuclei with larger number for wide range of $NV_{pn}$. TDA, which shows the constant straight line as a function of $NV_{pn}$, deviates both from RPA and the exact solution above approximately $NV_{pn}=0.2$. This result means that the ground state correlation resulted from the particle-particle interaction is important, as already mentioned in Ref. \cite{Stoica2001}.

\begin{figure}
\centering
\includegraphics[width=0.46\linewidth]{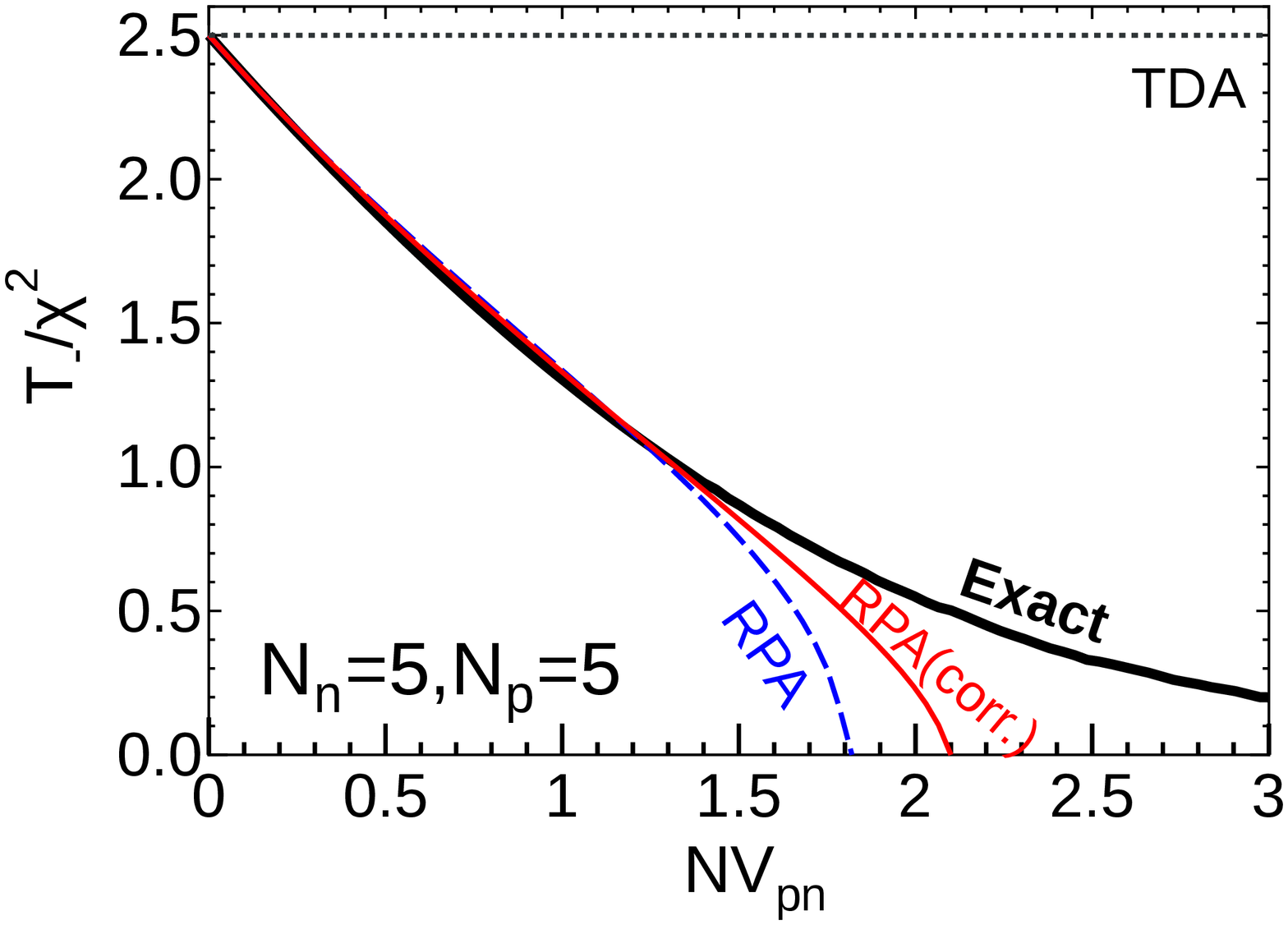}
\includegraphics[width=0.46\linewidth]{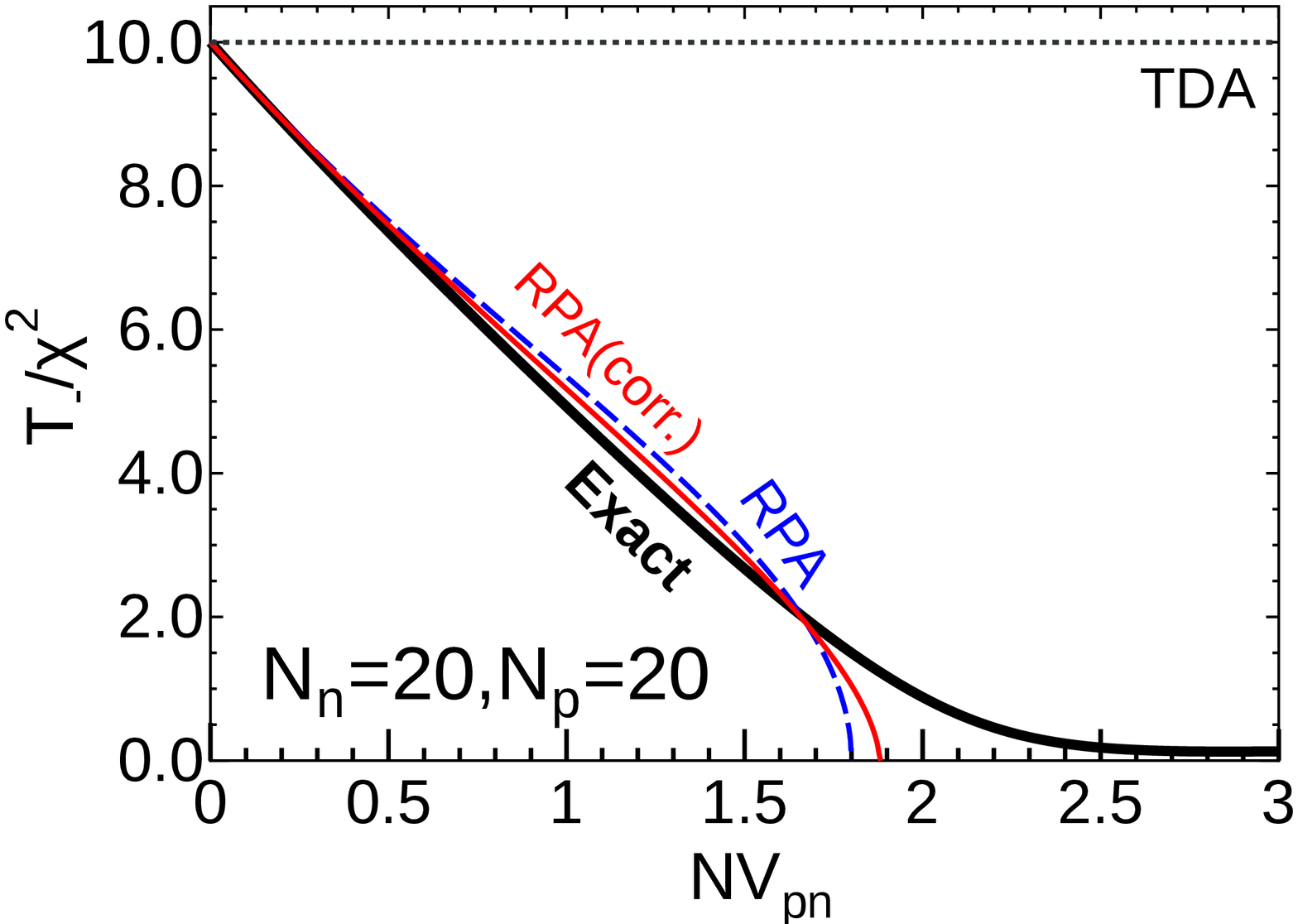}
\caption{Same as Fig. \ref{eNeqZ}, but for transition strengths.}
\label{tNeqZ}
\end{figure}

Figure \ref{tNeqZ} shows the transition strengths of the system of $N_n=N_p=5$ (the left panel) and $N_n=N_p=20$ (the right panel). 
Both RPA and RPA(corr.) show a similar result to the exact solution from $NV_{pn}=0$ to $\sim 1.5$. Above $NV_{pn}=1.5$, RPA collapses rapidly due to the phase transition. RPA(corr.) also collapses at a higher $NV_{pn}$ than RPA. The difference between them is, however, not as large as the excitation energies shown in Fig. \ref{eNeqZ}. Namely, the effect of the correlated ground state is not significant both for small and large nuclei for wide range of $NV_{pn}$. Again, TDA shows a large deviation from RPA and the exact solution, similar to the excitation energies.

\begin{figure}
\centering
\includegraphics[width=0.46\linewidth]{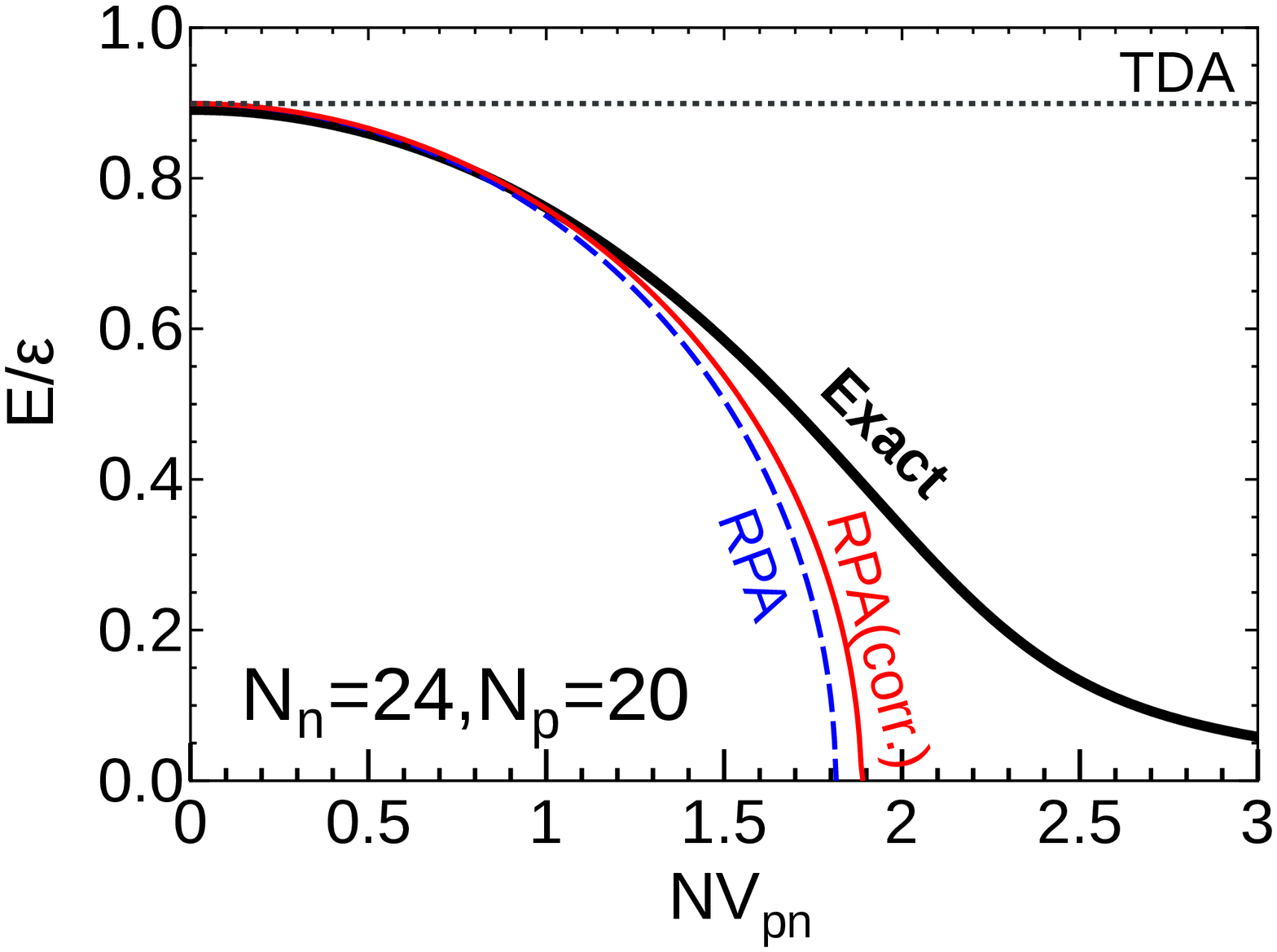}
\includegraphics[width=0.46\linewidth]{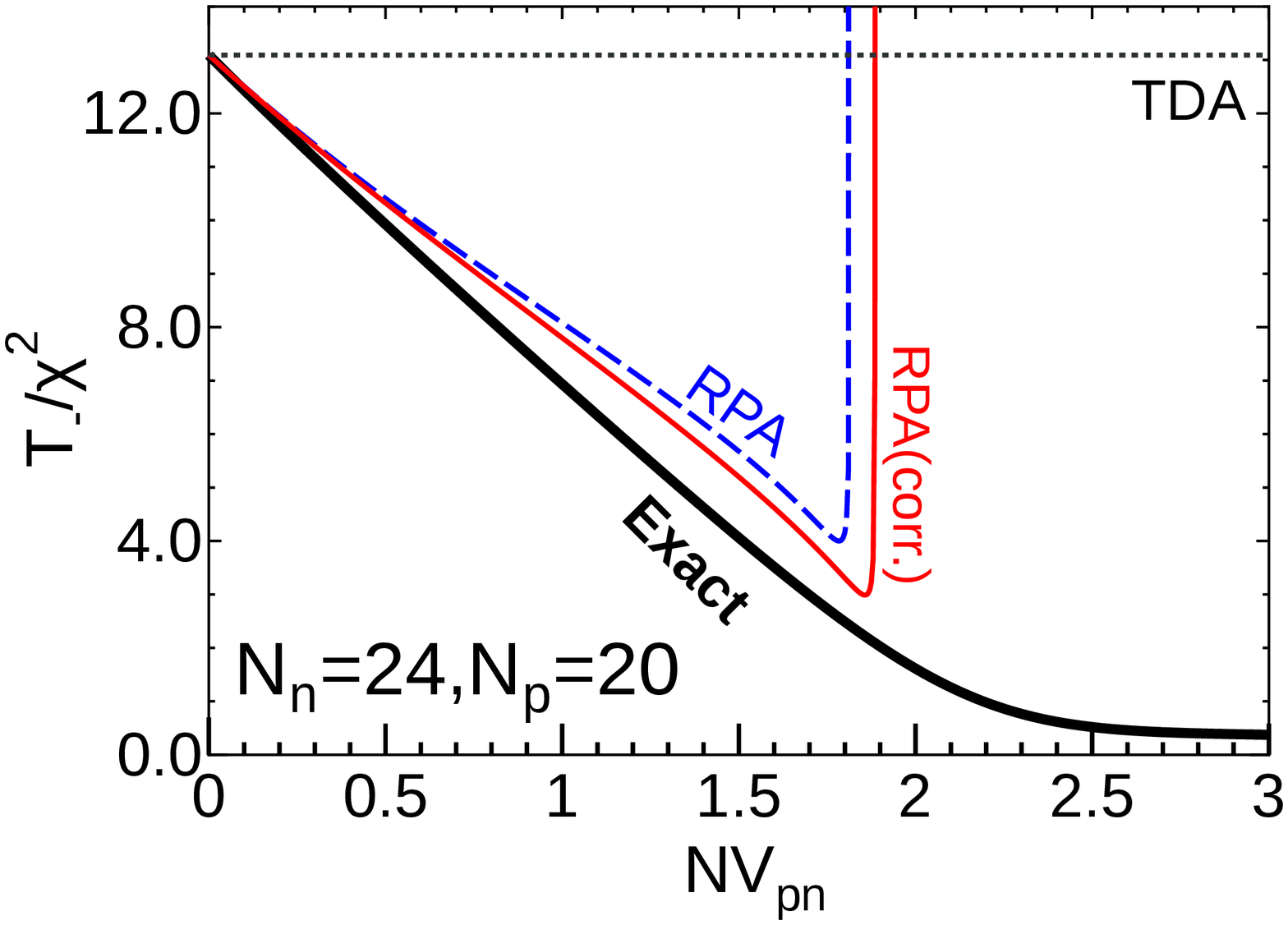}
\includegraphics[width=0.46\linewidth]{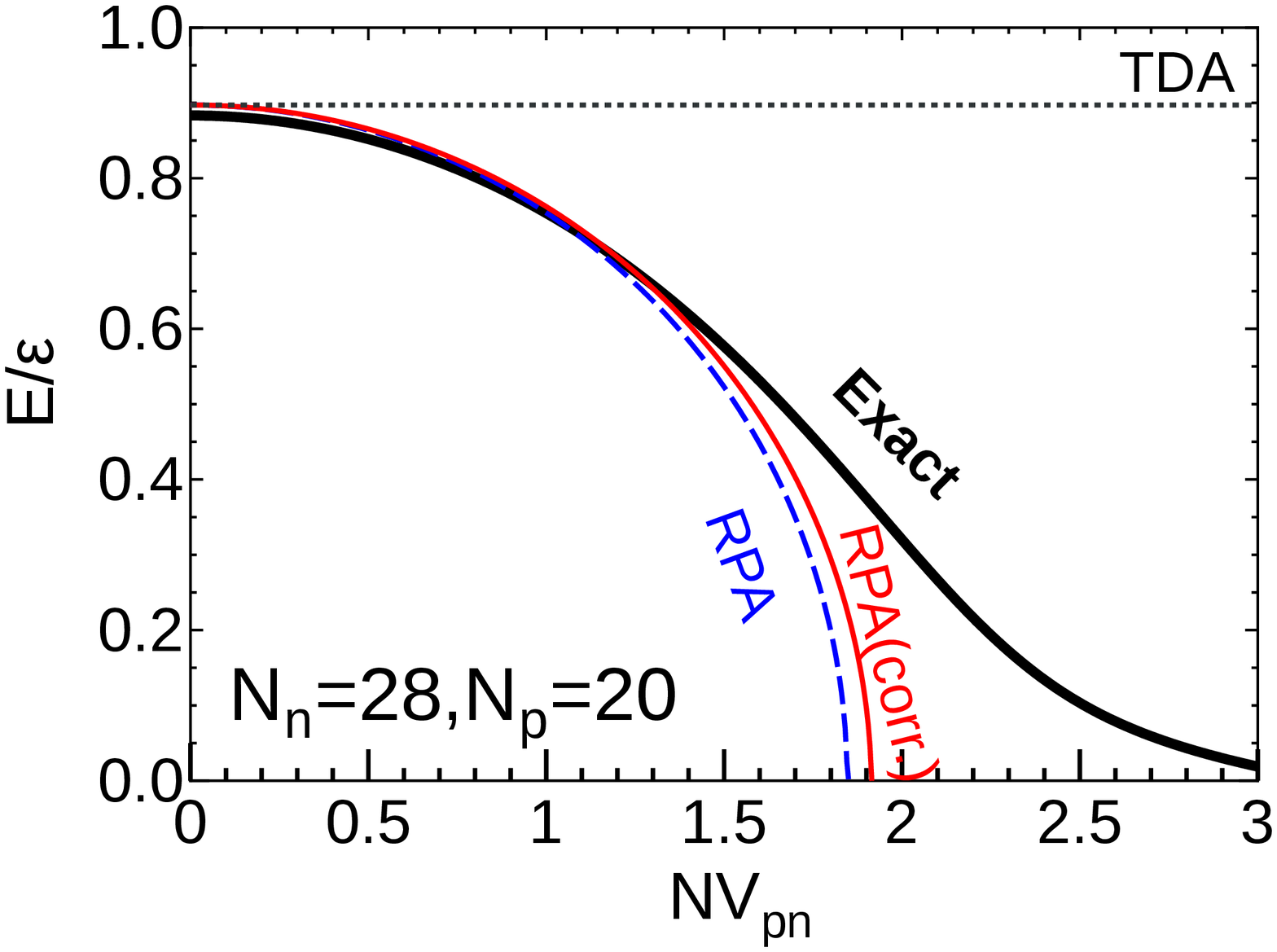}
\includegraphics[width=0.46\linewidth]{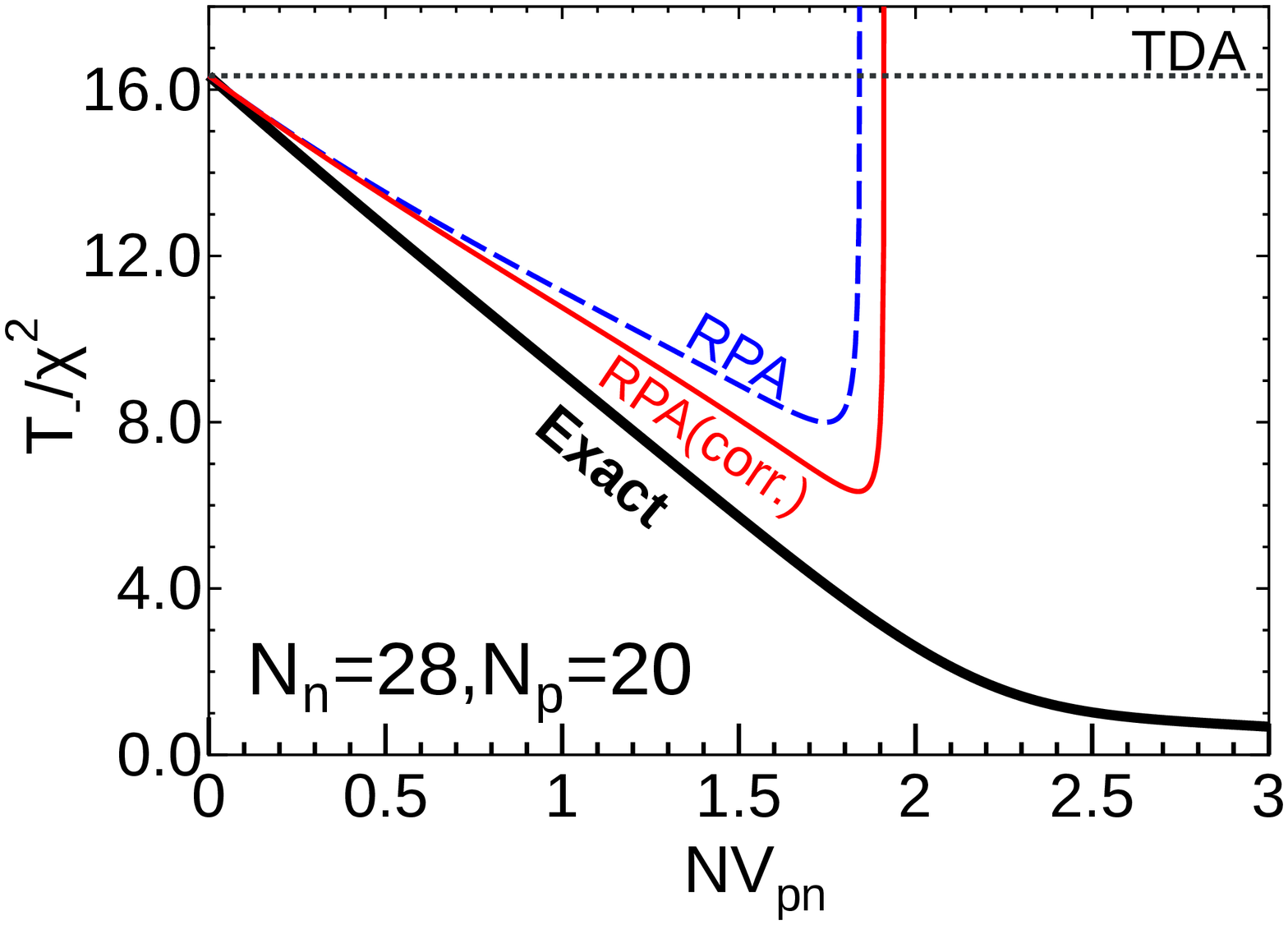}
\includegraphics[width=0.46\linewidth]{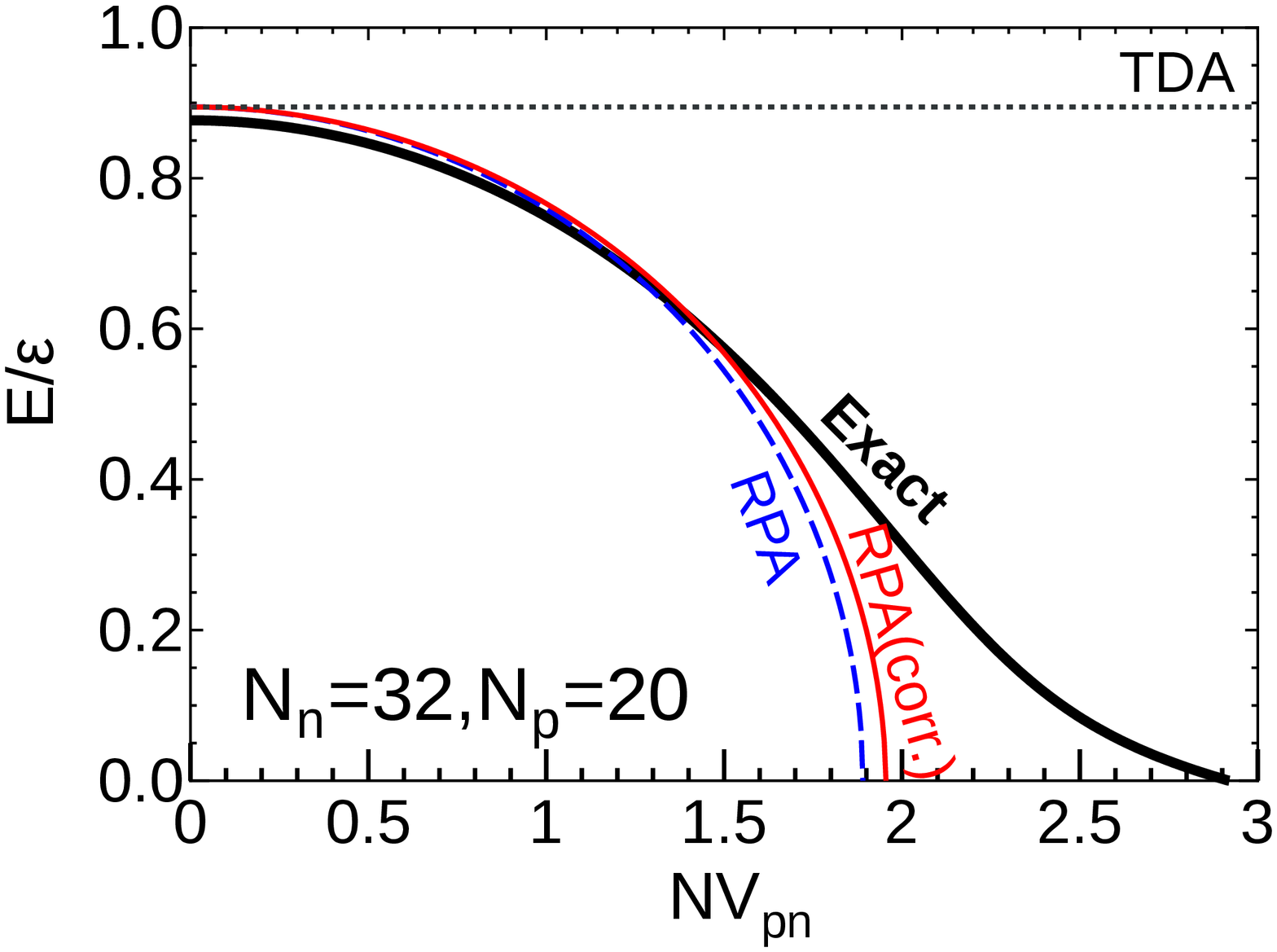}
\includegraphics[width=0.46\linewidth]{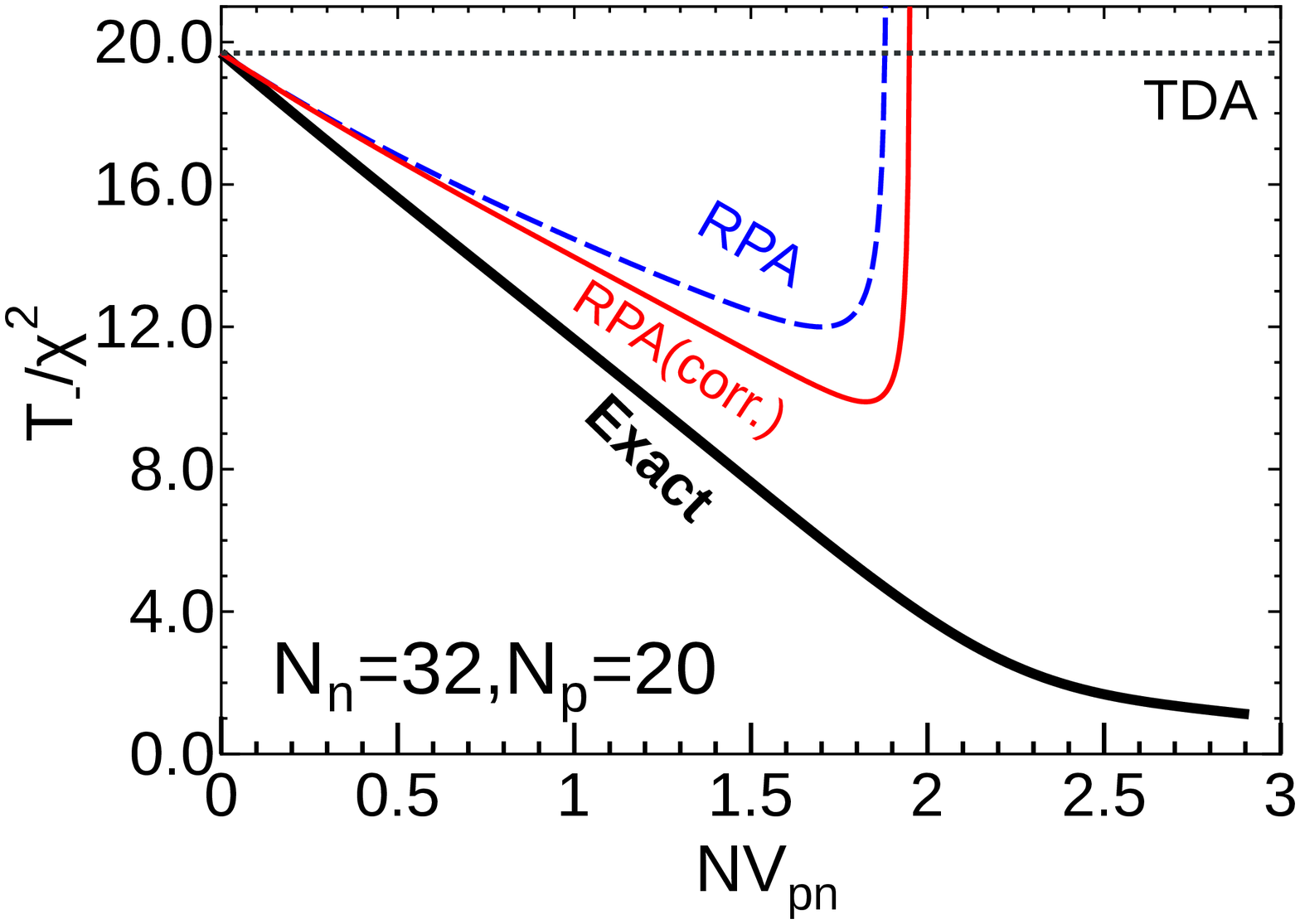}
\caption{Excitation energies (left panels) and transition strengths (right panels) in case of $N_n=24$ (top), $N_n=28$ (middle) and $N_n=32$ (bottom) as a function of $NV_{pn}$. The thick solid, thin solid, dashed, and dotted lines are the results for the exact, RPA(corr.), RPA, and TDA, respectively.}
\label{NneZ}
\end{figure}

Next we discuss the case of $N_n \ne N_p$. We keep $N_p=20$ and vary the neutron number from $N_n=24$ to $32$. The results are shown in Fig. \ref{NneZ}. The left and right panels illustrate the excitation energies and the transition strengths, respectively. The difference of the excitation energy between RPA, RPA(corr.) and the exact solution does not change significantly even if we change the $N_n$. The variations of the critical points of RPA and RPA(corr.) are also small between different $N_n$. However, the result of the transition strength shows a different tendency from the excitation energy. Looking at the right panels, the difference between RPA and the exact solution becomes larger as $N_n$ increases. The difference already starts at a small $NV_{pn}$. Let us remind that RPA and RPA(corr.) showed a good agreement with the exact solution in case of $N_n=N_p$ as seen in Fig. \ref{tNeqZ}. RPA(corr.) remedies the RPA result to some extent, but the difference from the exact one is still large.

\begin{figure}
\centering
\includegraphics[width=0.46\linewidth]{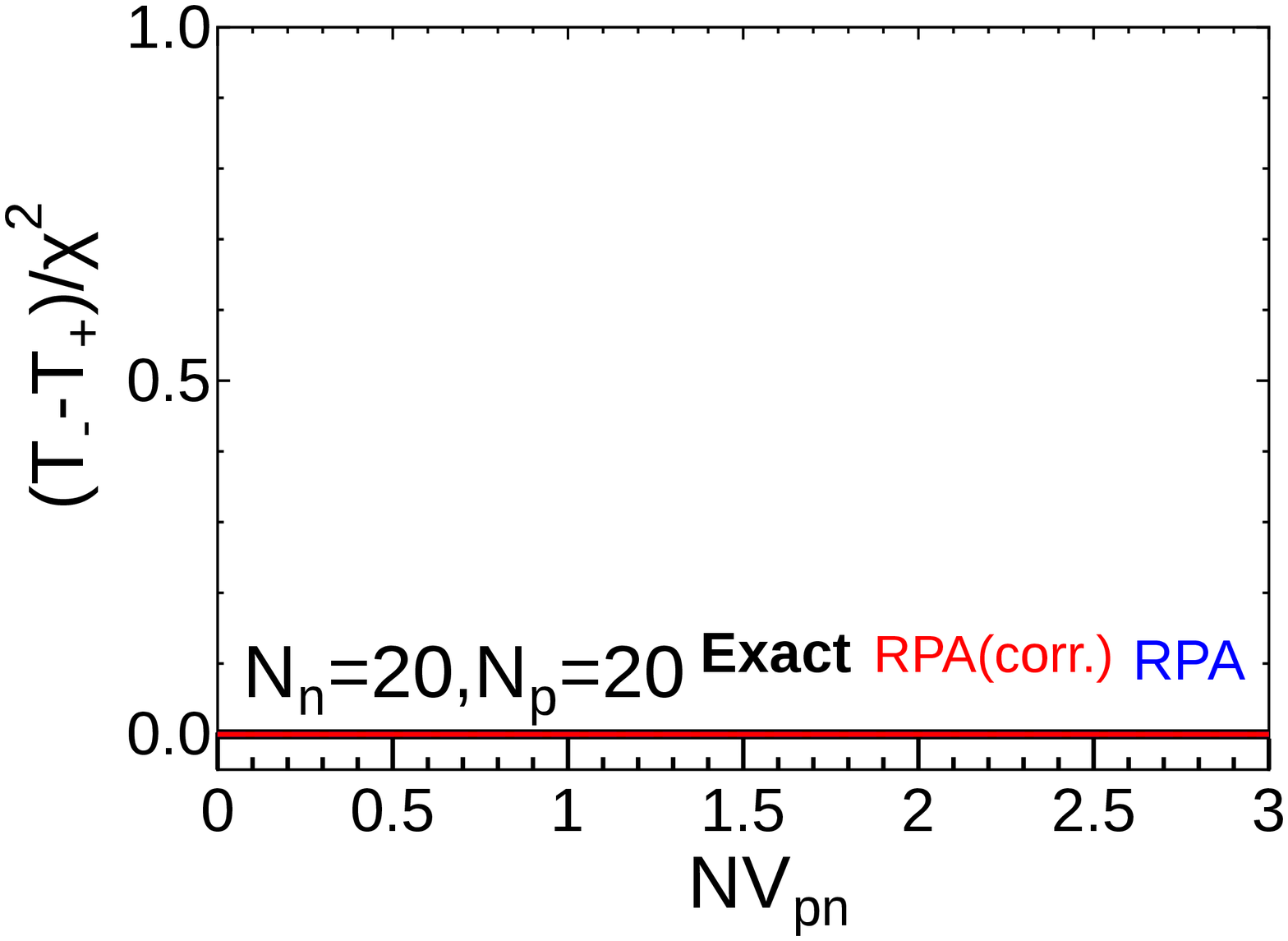}
\includegraphics[width=0.46\linewidth]{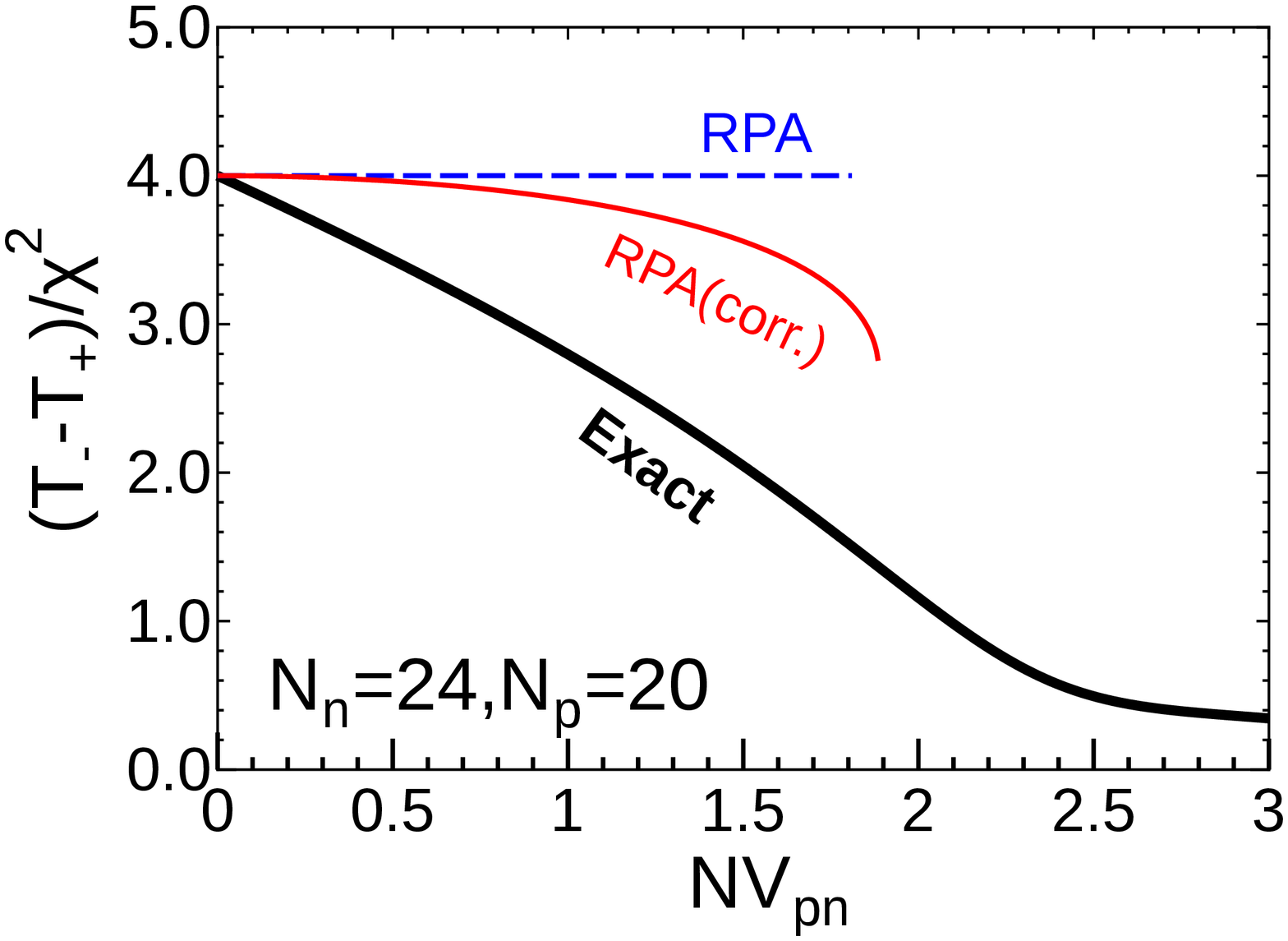}
\includegraphics[width=0.46\linewidth]{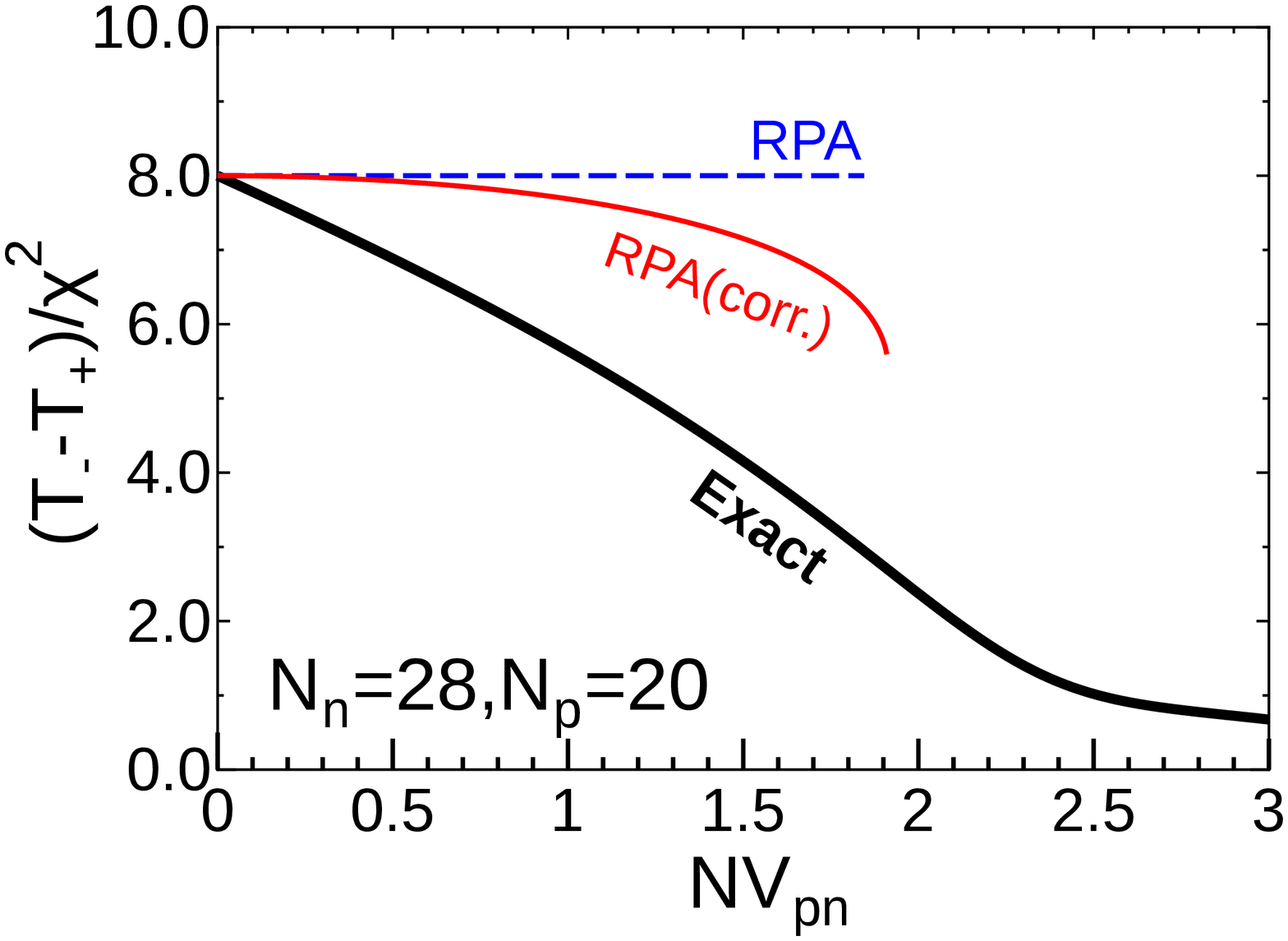}
\includegraphics[width=0.46\linewidth]{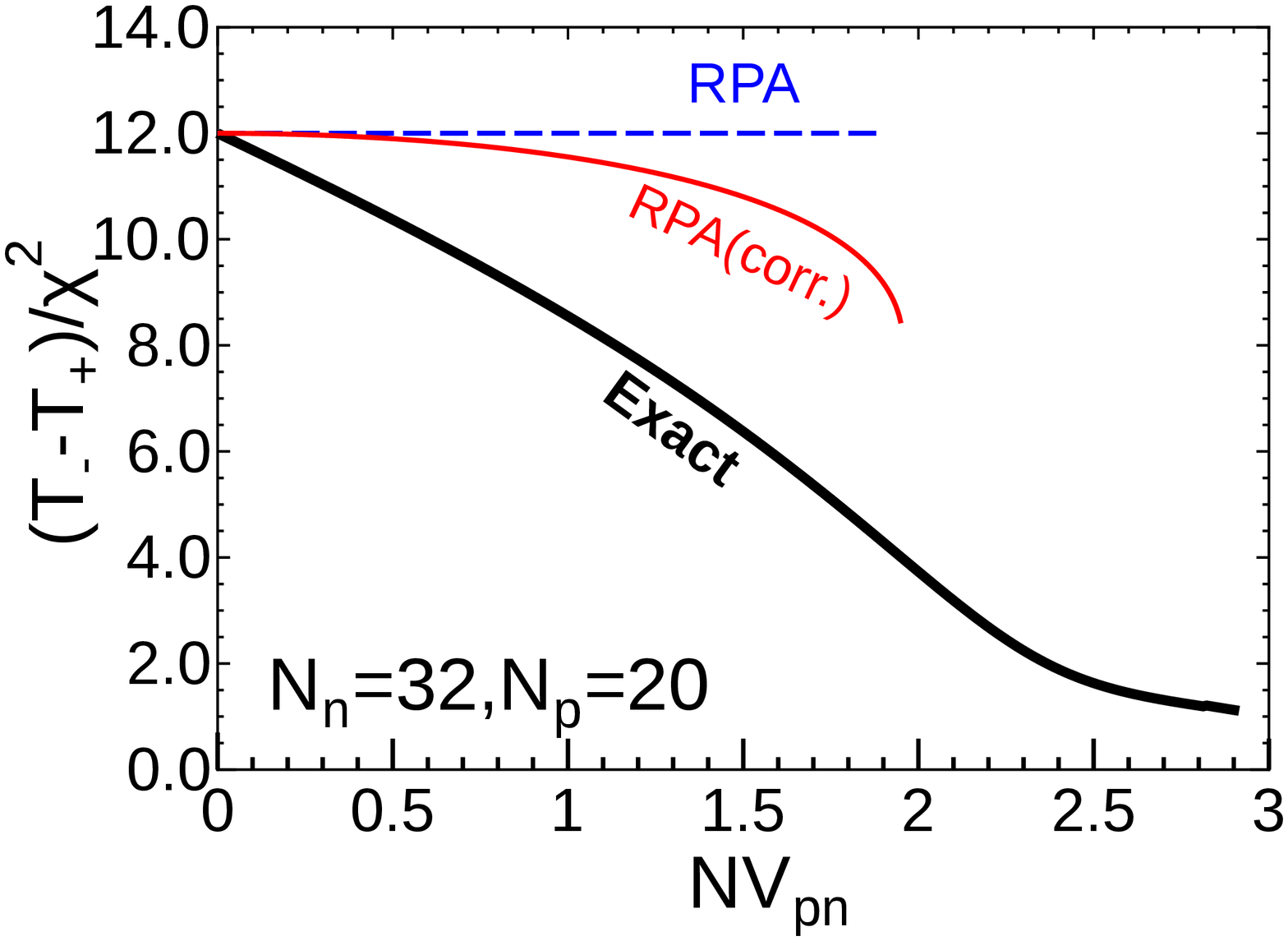}
\caption{Non energy weighted sum-rule, $T_--T_+$ for the first excited states. The thick solid, thin solid and dashed lines are the results for the exact, RPA(corr.), and RPA.}
\label{sumrule}
\end{figure}

We also investigate the non energy weighted sum-rule of charge exchange reaction defined by $T_--T_+$. The result is shown in Fig. \ref{sumrule}. While RPA perfectly satisfies the total sum-rule, which must be equal to $N_n-N_p$, up to the critical points, RPA(corr.) doesn't. The reason would be attributed from the fact that the phonon creation operator $\Gamma^\dagger$ doesn't consider the transition from the excited single particle states, as discussed in Ref. \cite{Gambacurta2006}. The exact solution also doesn't seem to satisfy the sum-rule. However, it satisfies the total sum-rule if we include the transition to other excited states besides the first one, which cannot be treated in RPA in two level model. It is clear that the difference between RPA and the exact solution becomes large when we consider the $N_n\ne N_p$ case. Analyzing the exact solution, transition to $2$ proton particle-$1$ proton hole $1$ neutron hole $[\pi^2(\pi\nu)^{-1}]$ configurations from the ground state becomes important, which cannot be connected by $M^+$ operator from the correlated ground state given by Eq. \eqref{RPAstate}. It is expected that the second RPA, which enables us to include such $2p2h$ configurations, can improve the result.

\section{Summary}
\label{summary}

We investigated the validity of charge exchange reaction using RPA with LMG model in SU(2)$\times$SU(2) basis. In case of which neutron and proton numbers are same, the RPA and RPA(corr.) works well both for small and large nuclei when the particle-particle interaction is weak. If the particle-particle interaction becomes strong, RPA and RPA(corr.) results begin to deviate from the exact solution. On the other hand, the transition strengths are still reproduced well. This situation changes in case of which neutron and proton numbers are different. The excitation energies are reproduced reasonably up to $NV_{pn}\sim1.5$, but the transition strengths are not. It turned out that the $2p2h$ configurations, which cannot be covered by the correlated ground state used in the present formalism, start to become important from a small $NV_{pn}$ value. It is expected that the extension of the model to include such a $2p2h$ configuration can reduce the difference between RPA and the exact solution. The work for it is now in progress.


\end{document}